\documentstyle[12pt]{article}
\begin{document}
\title{
\hfill{\small IPNO/TH 94-21}\\
\hfill{\small April 1994}\vspace*{0.5cm}\\
\sc Persistent current of free electrons in the plane\vspace*{0.5cm}\\
\normalsize (Comment on ``Relation between Persistent Currents
and\\
\normalsize the Scattering Matrix", Phys. Rev. Lett. {\bf 66}, 76 (1991))
\vspace*{0.3cm}}
\author{\sc Alain Comtet${}^1$,
Alexander Moroz\thanks{e-mail address :
{\tt moroz@ipncls.in2p3.fr}},\vspace{0.3cm}\\
\sc and St\'{e}phane Ouvry${}^1$\thanks{ouvry@ipncls.in2p3.fr}
}
\vspace{0.8cm}
\date{
\protect\normalsize
\it Division de Physique Th\'{e}orique\thanks{Unit\'{e} de Recherche des
Universit\'{e}s Paris XI et Paris VI
associ\'{e}e au CNRS}, IPN\\
\it Univ. Paris-Sud, F-91 406 Orsay Cedex\\
\it ${}^1$and LPTPE, Tour 12\\
\it Universit\'{e} Pierre et Marie Curie\\
\it 4 Place Jussieu, F-75 005 Paris\\
\it France
}
\maketitle
\begin{center}
{\large\sc abstract}
\end{center}
Predictions of Akkermans et al. are essentially changed when the Krein
spectral displacement operator is  regularized by means of zeta
function. Instead
of piecewise constant persistent current of free electrons on the plane
one has a current which varies linearly with the flux and is antisymmetric
with regard to all time preserving values of $\alpha$ including $1/2$.
Different self-adjoint extensions of the problem and role
of the resonance are discussed.

\vspace*{0.3cm}

\thispagestyle{empty}
\baselineskip 20pt
\newpage
\setcounter{page}{1}
\noindent
In \cite{AA} a persistent current around the origin,
pierced by a flux tube, for free electrons in the plane when all states
below the Fermi energy $E_F$  are occupied, was calculated.
The persistent current was defined with respect to a point. It was
given by the total current through a line that extends from that point
to infinity, in the absence of currents through the external leads.
Remarkably enough, a very compact formula
for the scattering-state contribution to the persistent current
was derived,
\begin{equation}
dI(E,\alpha) =(2\pi i)^{-1}\partial_\alpha[\ln\det\mbox{S}(E,\alpha)]dE,
\label{aaa}
\end{equation}
where $dI(E,\alpha)$ is the differential contribution to the
persistent current at energy $E$, $S(E,\alpha)$ is the on-shell
scattering matrix, and
$\alpha$ is the total flux $\Phi$ in the units
of the flux quantum $\Phi_o=hc/e$
\cite{AA}. We shall write $\alpha$ as $\alpha=n+\eta$
where $n=[\alpha]$ is the nearest  integer {\em smaller} than
$\alpha$, and $\eta$ is the nonintegral part of
 $\alpha$, $0\leq \eta<1$.

It is well known that the scattering
in the presence of the Aharonov-Bohm potential is {\em singular} since
the phase shifts $\delta_l(E)$ do not decay as the orbital
quantum number $l\rightarrow\infty$ \cite{AB}.
Therefore in \cite{AA}
$\lim_{L\rightarrow\infty}\sum_{|l|<L}\delta_l$
was used to define $\ln\det\mbox{S}$.
However, recently it has been shown \cite{AM} that when  $\ln\det\mbox{S}$
 is regularized by means of zeta function regularization
the Krein-Friedel formula \cite{F}
\begin{equation}
(2\pi i)^{-1}\ln\det\mbox{S}(E,\alpha)=\triangle N(E,\alpha),
\end{equation}
where $\triangle N(E,\alpha)$ is the change of the integrated
density of scattering states caused by the presence of a scatterer,
remains  also valid in the {\em singular} scattering problem.
The results for
$\triangle N(E,\alpha)$ for different self-adjoint
extensions of the problem can be read off from \cite{AM}.
The most general expression  for the contribution
of scattering states to $N_\alpha$ for $E\geq 0$
which includes the situation
 when  bound states are present and confirms calculations of
\cite{CGO} is
\begin{eqnarray}
\triangle N(E,\alpha) = &-&\frac{1}{2}\eta(1-\eta)+\frac{1}{\pi}\arctan
\left(\frac{\sin(\eta\pi)}{\cos(\eta\pi)-(|E_{-n}|/E)^{\eta}}
\right)
\nonumber\\
&-&\frac{1}{\pi}\arctan
\left(\frac{\sin(\eta\pi)}{\cos(\eta\pi)+
(|E_{-n-1}|/E)^{(1-\eta)}}\right),
\label{intsing}
\end{eqnarray}
where $E_{-n}$ and $E_{-n-1}$ are the binding energies
in $l=-n$ and $l=-1-n$ channels.
If the bound states are not present the {\em arctan} terms are missing
(the limit is achieved by letting bound state
energies go to $-\infty$ and not to zero).
The result (\ref{intsing}) for the particular self-adjoint extension
 without bound states was obtained originally in \cite{CGO} in
the context of anyonic physics.

Note that under the presence of bound states
one has for $0<\eta<1/2$ a {\em resonance} at
\begin{equation}
E=\frac{|E_{-n}|}{(\cos(\eta\pi))^{1/\eta}}>0
\label{reso}
\end{equation}
at $l=-n$-th channel. The phase shift
$\delta_{-n}(E)$ \cite{AM} changes by $\pi$ in the direction
of increasing energy and  the integrated density of states
(\ref{intsing}) has a sharp
increase by one. For $1/2<\eta<1$ the resonance is shifted to the
$l=-n-1$ channel.

As a result the predictions of \cite{AA} are essentially changed.
If (\ref{intsing}) (without the bound states)
 is inserted in (\ref{aaa}) one finds
\begin{equation}
I=-E_F(\eta-1/2),
\end{equation}
and the current becomes to depend linearly on $\eta$ (cf. \cite{AA}
where it is a constant).
As a bonus the current is also antisymmetric
with regard to $\eta=1/2$  which is the point
where the time invariance is preserved,
and it vanishes here
in contrast to \cite{AA}. Observations
of the persistent current may finally reveal the resonance
in the Aharonov-Bohm scattering which was predicted in \cite{AM},
since near it the current becomes very sensitive to the change of
the Fermi energy $E_F$ and of the flux near the value which satisfies
(\ref{reso}) for $E=E_F$.

A more elaborated discussion of the problem will be
given elsewhere \cite{CMO}.



\begin{thebibliography}{99}
\bibitem{AA}E. Akkermans, A. Auerbach, J. E. Avron, and B. Shapiro,
Phys. Rev. Lett.\ {\bf 66}, 76 (1991).
\bibitem{AB}
Y. Aharonov and D. Bohm, Phys. Rev.\ {\bf 115}, 485 (1959);
W. C. Henneberger, Phys. Rev. \ {\bf A22}, 1383 (1980).
\bibitem{AM}A. Moroz, IPNO/TH 94-20 (hep-th/9404104).
\bibitem{F}
J. M. Lifschitz, Usp. Matem. Nauk {\bf 7}, 170 (1952);
M. G. Krein, Matem. Sbornik {\bf 33}, 597 (1953);
J. S. Faulkner, J. Phys. C: Solid State Phys.\ {\bf 10}, 4661
(1977).
\bibitem{CGO}A. Comtet, Y. Georgelin, and S. Ouvry, J. Phys. A: Math. Gen.\
{\bf 22}, 3917 (1989).
\bibitem{CMO}A. Comtet, A. Moroz, and S. Ouvry, IPNO/TH 94-30
(in preparation).
\end{thebibliography}
\end{document}